\documentclass[amssymb,aps,twocolumn,floats,showpacs]{revtex4-1}
\usepackage{color,graphicx,pstricks,float}
\usepackage{natbib}

\usepackage{fancybox}
\usepackage{epsfig}
\usepackage{graphicx}
\usepackage{tabularx}
\usepackage{inputenc}
\usepackage{amsmath,braket,amsfonts,amsmath,mathtools,mathrsfs}
\usepackage{amssymb}
\usepackage{gensymb}
\usepackage{grffile}
\usepackage{relsize}
\usepackage{array}

\begin{document}

\title{Antiferromagnetic Skyrmions and Skyrmion Density Wave in Rashba Hund's Insulator}

\author{Arnob Mukherjee}
\author{Deepak S. Kathyat}
\author{Sanjeev Kumar}

\address{ Department of Physical Sciences,
Indian Institute of Science Education and Research Mohali, Sector 81, S.A.S. Nagar, Manauli PO 140306, India \\
}

\begin{abstract}
We discover magnetic phases hosting highly-elusive and technologically important antiferromagnetic skyrmion quasiparicles in a model for Rashba-coupled Hund's insulator. The results are based on unbiased simulations of a classical superexchange model derived, in this work, from a realistic microscopic electronic Hamiltonian. 
We also discover a novel skyrmion density wave groundstate characterized by a ($Q,Q$) modulation in the local skyrmion density map. A unique inhomogeneous state characterized by a circular pattern in spin structure factor and filamentary real-space textures is identified as the parent of sparse antiferromagnetic skyrmions. We predict that the magnetic states reported here can be realized in thin films of multiorbital systems involving $4d$ or $5d$ transition metals.
\end{abstract}
\date{\today}

\maketitle

\noindent
{\it Introduction:--}
Importance of skyrmion-like topological magnetization textures for spintronics applications is widely accepted \cite{AFert2017, Tokura2017, Jungwirth2016, Nagaosa2013b, Bogdanov2020}. However, the intrinsic skyrmion Hall effect -- the phenomenon of transverse deflection of these quasiparicles -- is a deterrent to their potential applications in race-track memory devices. It has been demonstrated that the skyrmion Hall effect gets strongly suppressed for an antiferromagnetic skyrmion (ASk), which is a configurational linear combination of an antiferromagnet and a skyrmion \cite{Legrand2020, Zhang2016, Barker2016, Kim2017, Akosa2018a, Woo2018a, Gomonay2018, Dohi2019}. 
%Such ASk textures have been recently reported in experiments on FeGdCo thin films \cite{Woo2018a}. 
While a strategy to engineer ASk textures has been proposed in bilayer systems \cite{Gobel2017, Bode2006}, general guiding principles based on microscopic model Hamiltonians for identifying candidate materials are currently unavailable.
%Theoretical understanding of ASk textures is mostly based on micromagnetic simulations and phenomenological spin models. Microscopic understanding of formation of ASk is currently lacking, and therefore the search for ASk textures is currently based on approximate guiding principles, such as the presence of Dzyaloshinskii-Moriya (DM) coupling. 

Recent studies have shown that the Hund's coupling plays a key role in determining the nature of transport and magnetism in correlated multi-band systems \cite{Georges2013, DeMedici2011, DeMedici2017, Isidori2019}. The interplay between repulsive Hubbard interactions, and Hund's coupling has given rise to new concepts, such as, Hund's metal, Hund's insulator, Mott-Hund's insulator and topological Hund's insulator \cite{Fanfarillo2015, Mcnally2015, Budich2013}. Some of the materials that highlight the unusual physics emerging from Hund's coupling are, LaMnO$_3$, BaMn$_2$As$_2$, and LaCrO$_3$ \cite{Mcnally2015, Millis1997}. Ir, Ru, Os and Tc based oxides further enrich this interplay by bringing in a moderate to strong spin-orbit coupling \cite{Watanabe2010a, Meetei2015, Sutter2017, Paramekanti2018}. The influence of Rashba SOC on magnetism of Hund's insulators is an outstanding theoretical problem with potentially significant implications for skyrmion and ASk physics.
%A question of fundamental importance is whether Hund's physics can serve as a proxy for Mott physics for understanding the effect of SOC on the magnetic ordering in strongly correlated systems?

In this work, we derive and investigate a classical superexchange Hamiltonian arising from a Rashba coupled Hund's insulator. We uncover a rich magnetic phase diagram consisting of unconventional ordered as well as disordered phases. Most notable are the liquid-like antiferromagnet string (AFS) state and the ASk lattice (ASkL) state. In the presence of external magnetic field, we find a phase hosting sparse ASk (s-ASk) textures. For strong Rashba SOC, we unveil novel skyrmion density wave (SkDW) state that generalizes the concept of ferromagnetic and antiferromagnetic skyrmion lattices. Our study shows that the interplay of Hund's coupling with Rashba SOC generates the classical version of the physics expected in Rashba-Mott insulators \cite{Farrell2014a, Banerjee2014}. Being less susceptible to quantum fluctuations, materials that realize the Rashba-Hund's mechanism are strong candidates for providing stable ASk textures.

\noindent
{\it Classical Superexchange Hamiltonian:--}
The ferromagnetic Kondo lattice model (FKLM) in the presence of Rashba SOC on a square lattice is described by the Hamiltonian,

\begin{eqnarray}
H & = & - t \sum_{\langle ij \rangle,\sigma} (c^\dagger_{i\sigma} c^{}_{j\sigma} + {\textrm H.c.}) 
+ \lambda \sum_{i} [(c^{\dagger}_{i \downarrow} c^{}_{i+x\uparrow} - c^{\dagger}_{i\uparrow} c^{}_{i+x\downarrow}) \nonumber \\
& & + \textrm{i} (c^{\dagger}_{i\downarrow} c^{}_{i+y\uparrow} + c^{\dagger}_{i\uparrow} c^{}_{i+y\downarrow}) + {\textrm H.c.}] - J_{\textrm H} \sum_{i} {\bf S}_i \cdot {\bf s}_i.
\label{Ham}
\end{eqnarray}

\noindent
The annihilation (creation) operators, $c_{i\sigma}$ ($c_{i\sigma}^\dagger$), satisfy the usual fermion algebra.
$J_{\rm H}$ ($\lambda$) denotes the strength of Hund's (Rashba) coupling, $t$ is the nearest neighbor hopping parameter.
$\bf{s}_i$ and ${\bf S}_i$ denote, respectively, the electronic spin operator and localized classical spin at site $i$. 
We parameterize by $\alpha$ the strength of hopping amplitude and Rashba coupling as $t = (1-\alpha) t_0$ and $\lambda = \alpha t_0$, where $t_0=1$ sets the reference energy scale. For large $J_{\textrm H}$, the physics of the model can be well described by the, $J_{\textrm H} \rightarrow \infty$, double-exchange limit, provided the electronic filling fraction is different from half \cite{Kathyat2020a}. Here, we focus on the half-filled case and derive the classical superexchange model via second order perturbation theory \cite{SM}.
The resulting spin Hamiltonian is given by,

\begin{eqnarray}
H_{\textrm{CSE}} &=& -1/J_{\textrm H} \sum_{i,\gamma} \big[ t^2(1-{\bf S}_i \cdot {\bf S}_j) - 2t\lambda  \hat{\gamma'} \cdot ({\bf S}_i \times{\bf S}_j)  \nonumber  \\
& & + \lambda^2(1+{\bf S}_i \cdot {\bf S}_j-2 (\hat{\gamma'} \cdot {\bf S}_i)(\hat{\gamma'} \cdot {\bf S}_j)) \big],
\label{eq:ESH}
\end{eqnarray}

\noindent
where site $j$ is the nearest neighbor (nn) of $i$ in $+{\hat{\gamma}}$ direction with $\gamma \in \{x,y\}$ and $\hat{\gamma'} = \hat{z} \times \hat{\gamma}$. Note that the rigorous derivation of the classical model is much simpler than that of the quantum model for Mott insulators via Schrieffer-Wolff transformation \cite{Schrieffer1966}. However, promoting the classical spin variables to spin operators leads to the identical quantum spin model. It will be interesting to test the generality of this classical-quantum correspondence for derivation of Kugel-Khomskii spin-orbital models by constructing toy classical Hamiltonians \cite{Kugel1982}. We have explicitly checked the validity of the model by comparing energies of various phases between the original quantum Hamiltonian Eq. (\ref{Ham}) and the $H_{{\rm CSE}}$ Eq. (\ref{eq:ESH}) \cite{SM}.
%\begin{eqnarray}
%f^{y}_{ij} = \frac{1}{2} \bigg[t^2(1-{\bf S}_i. {\bf S}_j) + \lambda^2(1+{\bf S}_i. {\bf S}_j-2S_i^{x}S_j^{x}) + 2t\lambda({\bf S}_i \times{\bf S}_j)_x]
%\label{eq.10}
%\end{eqnarray}

%\begin{figure}[t!]
%\includegraphics[width=.92 \columnwidth,angle=0,clip=true]{fig1.pdf}
%\caption{(Color online) (a) - (b) Magnetization $M_z$ (triangles), total skyrmion density $\chi$ (circles) and helicity $\eta$ (squares) as a function of applied Zeeman field for different values of Dresselhaus strength $\alpha$. Left $y$-axis scale is for TSD.
%}
%\label{fig1}
%\end{figure}

%\noindent
%{\it Energy comparison between exact and derived models:--} 

\noindent
{\it Ground state phase diagram:--} The presence of various competing anisotropic terms makes it difficult to analytically identify the ground states of $H_{\textrm{CSE}}$.
Nevertheless, it is instructive to note certain general features of the model in the limiting cases. For $\alpha = 0$, the model reduces to an isotropic Heisenberg model with antiferromagnetic ground state.  For $\alpha = 0.5$, the model becomes maximally anisotropic since the isotropic parts in the first and the third terms exactly cancel. Large $\alpha$ limit supports a kitaev-like anisotropic form and should lead to degenerate ground states. In order to investigate the magnetic phases of $H_{\textrm{CSE}}$, we perform large scale classical Monte Carlo simulations.
The different magnetic phases are characterized via component resolved SSF,

\begin{eqnarray}
S^{\mu}_{f}({\bf q}) &=& \frac{1}{N^2}\sum_{ij} S^{\mu}_i S^{\mu}_j~ e^{-{\rm i}{\bf q} \cdot ({\bf r}_i - {\bf r}_j)},
\label{SSF}
\end{eqnarray}

\noindent
where, $\mu = x, y, z$ denotes the component of the spin vector and ${\bf r}_i$ is the position vector for site $i$. The total structure factor can be computed as, 
$S_f({\bf q}) = \sum_{\mu} S^{\mu}_{f}({\bf q})$. Averaging over Monte Carlo steps is implicitly assumed.

\begin{figure}[t!]
\includegraphics[width=.86 \columnwidth,angle=0,clip=true]{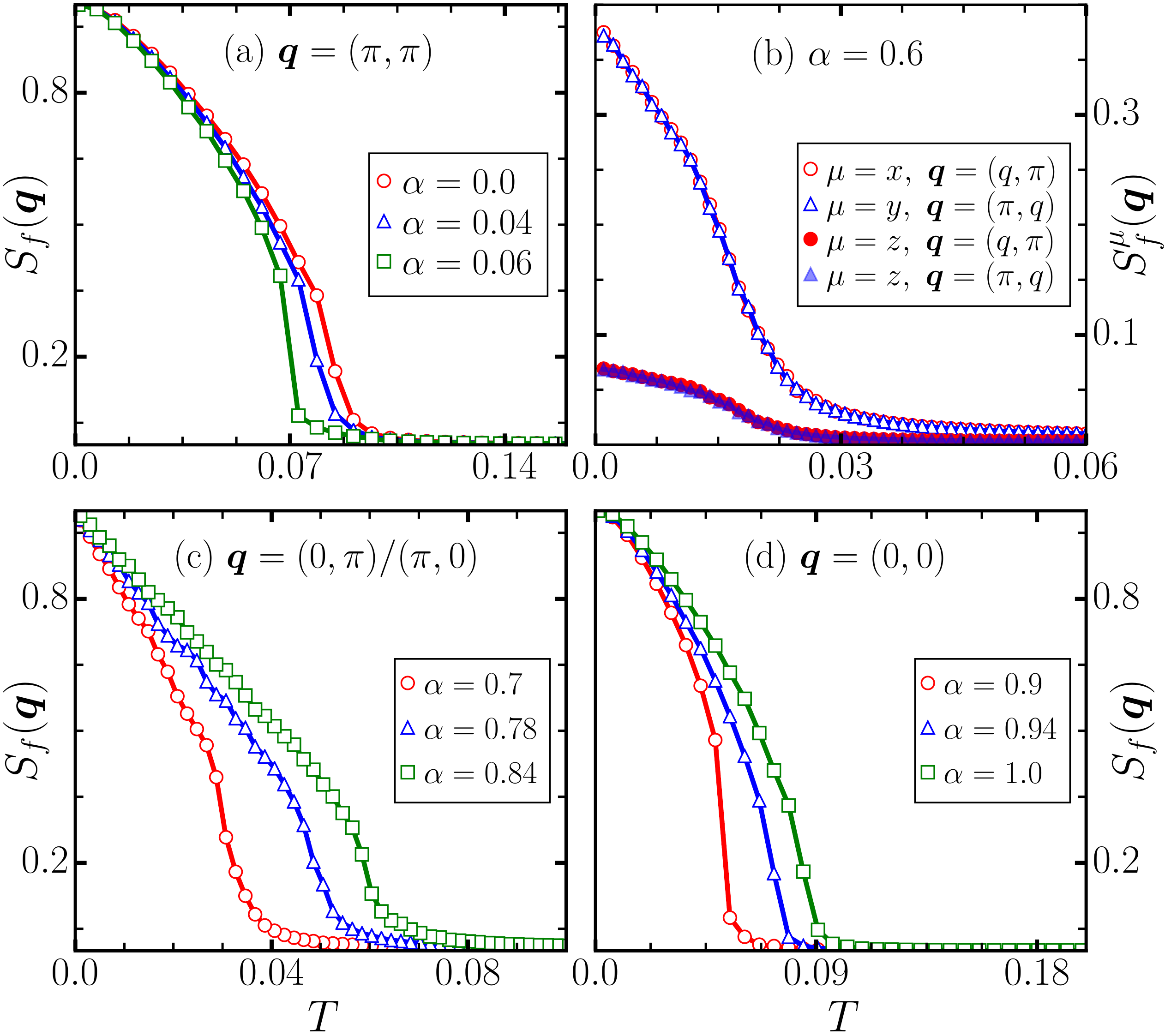}
\caption{(Color online) (a)-(d) Different components of SSF as a function of temperature for different $\alpha$. Panel (b) shows different components that are simultaneously finite, whereas panel (c) indicates that either ($0,\pi$) or ($\pi, 0$) are finite.
}
\label{fig1}
\end{figure}

\begin{figure}[t!]
\includegraphics[width=.9 \columnwidth,angle=0,clip=true]{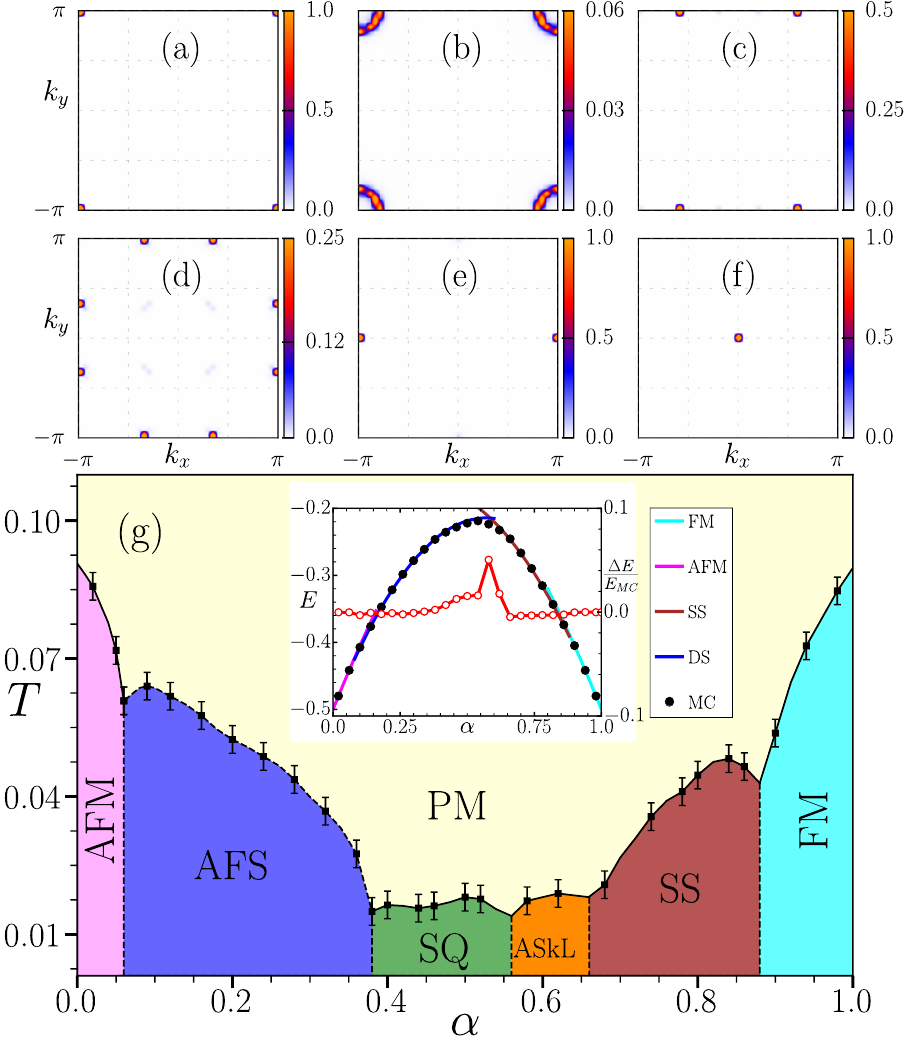}
\caption{(Color online) SSF at $T = 0.001$ for different values of $\alpha$ characterizing, (a) antiferromagnetic ($\alpha = 0$), (b) antiferromagnetic strings ($\alpha = 0.24$), (c) single-Q ($\alpha = 0.44$), (d) noncoplanar antiferromagnetic skyrmion lattice ($\alpha = 0.6$), (e) spin stripe ($\alpha = 0.7$), and (f) ferromagnetic ground states ($\alpha = 1$). (g) Phase diagram in the $T-\alpha$ plane. Inset in (g) compares the energy per site obtained via Monte Carlo simulations and that via variational calculations. Right $y$-axis displays the fractional difference between the two. DS refers to degenerate spiral states \cite{SM}.
}
\label{fig2}
\end{figure}

For small $\alpha$, we find a characteristic rise in the ($\pi,\pi$) component of the SSF upon reducing temperature (see Fig. \ref{fig1}(a)) suggestive of a transition to an antiferromagnetic ground state. For $\alpha = 0.6$, multiple components at multiple ${\bf q}$ points display order parameter like increase upon lowering temperature, as shown in Fig. \ref{fig1}(b). As we will show later, the noncoplanar ground state in this $\alpha$-regime is, in fact, an ASkL. For large $\alpha$, we find a spin-stripe state with peak at either ($0,\pi$) or ($\pi, 0$) in the SSF (see Fig. \ref{fig1}(c)), and finally a FM state (see Fig. \ref{fig1}(d)).

In Fig. \ref{fig2}(a)-(f), we present the evolution of magnetic ground states, as represented by low temperature SSF, from antiferromagnet to FM. SSF peaks at ($\pi,\pi$) for $\alpha < 0.06$ (see Fig. \ref{fig2}(a)).
Over a wide range of $\alpha$ values, $0.06 \leq \alpha \leq 0.38$,  we find that the SSF does not display any prominent peaks, but instead a diffuse ring pattern centered at ($\pi,\pi$) (see Fig. \ref{fig2}(b)). This indicates the presence of a disordered state that we will discuss later in detail. 
The circular pattern centered at ($\pi,\pi$) is similar to that reported in the Rashba double exchange model with center at ($0,0$) \cite{Kathyat2020a}.
Upon increasing $\alpha$ further, the state with circular pattern evolves into a doubly-degenerate state with SSF peaks at either ($Q,\pi$) or ($\pi, Q$) (see Fig. \ref{fig2}(c)). We label this as a single-Q (SQ) state. Fig. \ref{fig2}(d) displays a multi-Q pattern of peaks in SSF, typical of noncoplanar states. A doubly-degenerate spin stripe state, characterized by SSF peak at ($0,\pi$) or ($\pi, 0$) is stabilized for larger $\alpha$ (see Fig. \ref{fig2}(e)). Finally a FM state becomes the ground state as confirmed by a ($0,0$) peak in SSF shown in Fig. \ref{fig2}(f).

Using the temperature dependence of the characteristic features in SSF as order parameters for different magnetic phases, we summarize our findings in the form of a magnetic phase diagram in $T-\alpha$ plane shown in Fig. \ref{fig2}(g). Inset in Fig. \ref{fig2}(g) compares the ground state energies between the simulations and the variational calculations. The energies match very well, except for the ASkL states where no simple variational ansatz captures the important features of the ground states obtained via Monte Carlo. The variational analysis is particularly useful in uncovering degeneracies that remain hidden in the Monte Carlo simulations \cite{SM}. The antiferromagnet string states can be considered as antiferromagnetic analog of the liquid-like fDW states reported recently in the generalized double exchange model \cite{Kathyat2020a}. 

\begin{figure}[t!]
\includegraphics[width=.90 \columnwidth,angle=0,clip=true]{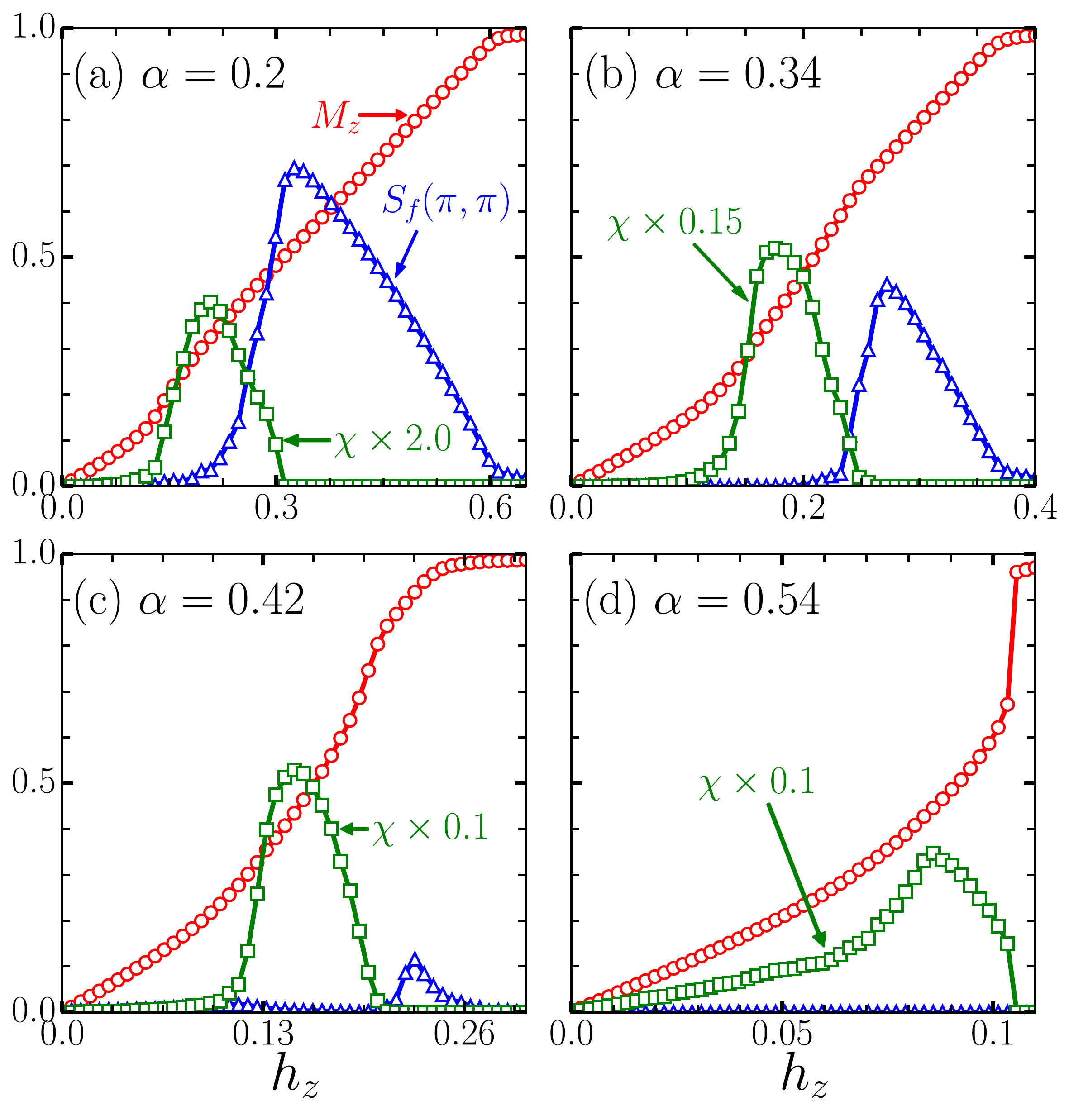}
\caption{(Color online) (a)-(d) Variations of magnetization, staggered magnetization and skyrmion density with Zeeman field for different values of $\alpha$.
}
\label{fig3}
\end{figure}

\noindent
{\it Antiferromagnetic skyrmions and skyrmion density wave:--} 
Having established that the SE model supports a rich variety of magnetic phases, we investigate the effect of applied magnetic field on these unusual ground states. We add a Zeeman term, $-h_z \sum_i S^z_i$, to the Hamiltonian Eq. (\ref{eq:ESH}) with $h_z$ denoting the strength of applied magnetic field. The evolution of the magnetic ground states with increasing $h_z$ is studied via Monte Carlo simulations with the zero-field cooled (ZFC) protocol. In addition to the SSF, we calculate local skyrmion density, defined as \cite{Chen2016},
\begin{eqnarray} 
\chi_{i} & = & \frac{1}{8\pi} [ {\bf S}_i \cdot ({\bf S}_{i+x} \times {\bf S}_{i+y} ) + {\bf S}_i \cdot ({\bf S}_{i-x} \times {\bf S}_{i-y})],
\end{eqnarray}
\noindent
and the skyrmion density structure factor ($\chi$SF). Upon increasing $h_z$, the magnetization, $M_z = \frac{1}{N} \sum_i S^z_i$, shows the expected rise originating from a continuous canting of spins towards the field direction (circle symbols in Fig. \ref{fig3}(a)-(d)). The deviation from this linear rise of $M_z$ is accompanied by a rise of the total skyrmion density, $\chi = \sum \chi_i$ (square symbols in Fig. \ref{fig3}(a)-(d)). The decrease in $\chi$ at larger $h_z$ is accompanied by an increase in the ($\pi,\pi$) component of SSF (triangle symbols in Fig. \ref{fig3}(a)-(c)). This confirms the presence of a canted antiferromagnet (CAF) state which continuously evolves towards a saturated FM phase. The variations with $h_z$ of the magnetization, staggered magnetization and the skyrmion density allows us to identify different phases in the model. We find that in some regimes of the phase space, the structure factor of $\chi_i$ is required to characterize the ordered states.

\begin{figure}[t!]
\includegraphics[width=.90 \columnwidth,angle=0,clip=true]{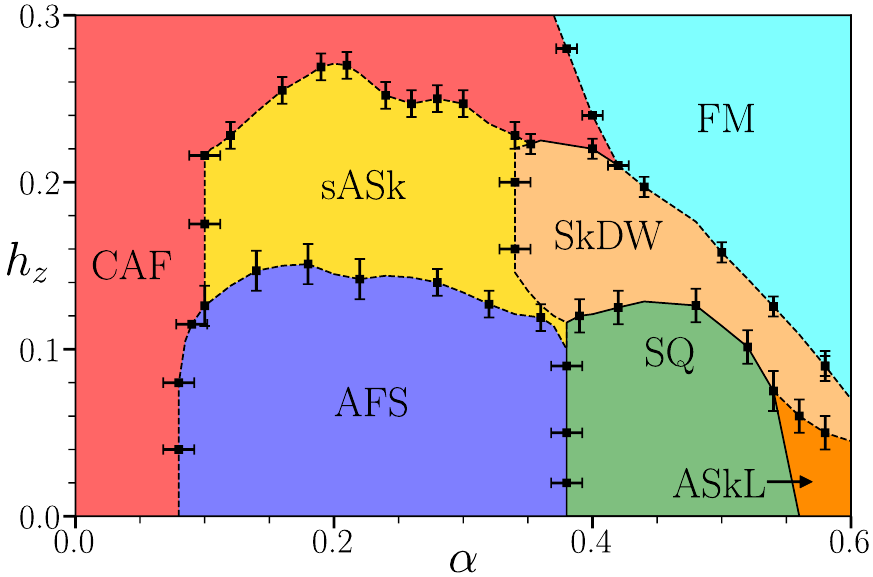}
\caption{(Color online) Low temperature phase diagram in $h_z$-$\alpha$ plane. The different phases are: (i) canted antiferromagnet (CAF), (ii) antiferromagnetic strings (AFS), (iii) sparse antiferromagnetic skyrmions (sASk), (iv) single-Q spiral (SQ), (v) skyrmion density wave (SkDW), (vi) antiferro skyrmion lattice (ASkL), and (vii) saturated ferromagnet (FM). The boundaries are inferred from the order parameters shown in Fig. \ref{fig3}, and the evolution of real-space maps shown in Fig. \ref{fig5}.
}
\label{fig4}
\end{figure}

\begin{figure*}[t!]
\includegraphics[width=1.88 \columnwidth,angle=0,clip=true]{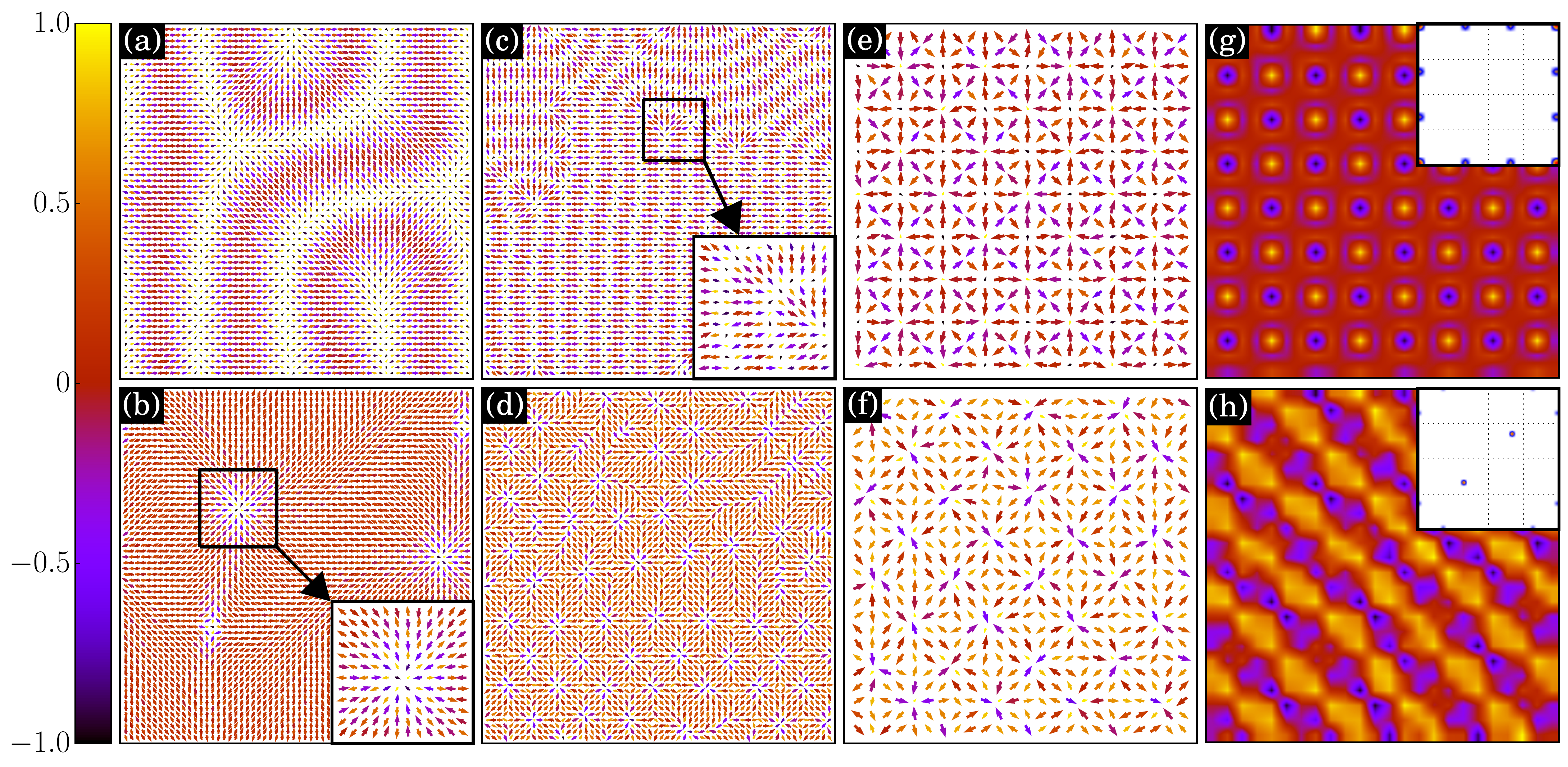}
\caption{(Color online) (a)-(f) Spin configuration snapshots in different magnetic phases: (a) AFS at $\alpha = 0.1$, $h_z = 0.0$, (b) sASk at $\alpha = 0.1$, $h_z = 0.15$, (c) AFS at $\alpha = 0.3$, $h_z = 0$, (d) sASk at $\alpha = 0.3$, $h_z = 0.17$, (e) ASkL at $\alpha = 0.6$, $h_z = 0$, and (f) SkDW at $\alpha = 0.4$, $h_z = 0.16$. The color map indicates the value of $S^z_i$ while the $x$ and $y$ components of spin are represented by arrows.
(g) and (h) display the skyrmion density ($\chi_i$) map for configurations shown in panels (e) and (f), respectively. The insets in (g) and (h) display the peak locations of the skyrmion density structure factor. 
}
\label{fig5}
\end{figure*}

Fig. \ref{fig4} presents a summary of our simulation results in the form of a ground state phase diagram in $\alpha$-$h_z$ plane. 
For small values of $\alpha$, the staggered AFM state at $h_z = 0$ continuously transforms into a CAF state and approaches the saturated FM state for large $h_z$. For intermediate $\alpha$ range, the AFS state at $h_z = 0$ evolves into a state with finite $\chi$ and then, at larger $h_z$, into CAF state. The real-space configurations confirm that the state with finite $\chi$, in fact, hosts isolated ASk textures. The real-space plots for all unusual states will be discussed below in detail. 
For $\alpha$ values that support SQ state in the absence of magnetic field, a novel state emerges beyond a critical applied field. This state can be described as a skyrmion density wave (SkDW). The SkDW state is also obtained by applying Zeeman field on the ASkL. For $\alpha > 0.42$, the CAF state ceases to exist and the system directly enters the FM phase from the SkDW. 

In Fig. \ref{fig5}, we present a microscopic view of the different phases described in the phase diagrams in Fig. \ref{fig2}(g) and Fig. \ref{fig4}.
The low-temperature spin configurations are shown in Fig. \ref{fig5}(a)-(f) for different phases as specified in the caption. In the absence of external field, the inclusion of Rashba SOC leads to an unusual fragmentation of the antiferromagnetic state with AF regions restricted to narrow filamentary channels that are separated by domain walls (see Fig. \ref{fig5}(a)). Interestingly, the filaments have an energetic freedom, originating from an unusual degeneracy in the model, to orient along any direction \cite{SM}.
For larger values of $\alpha$, the channels as well as the domain walls become narrow (see Fig. \ref{fig5}(c)).
On application of external field, these filamentary states give rise to sASk textures (see Fig. \ref{fig5}(b), (d)). For strong SOC, the $h_z = 0$ state already leads to a regular non-coplanar spin pattern (see Fig. \ref{fig5}(e)). In order to better understand the underlying structure of this spin configuration, we plot the derived scalar field, $\chi_i$, for the same state. We find an unusual antiferro pattern of local skyrmion density (see Fig. \ref{fig5}(g)). The corresponding $\chi$SF displays, in addition to peaks located at ($\pi$,$\pi$), large intensity at Brillouin zone edges similar to that obtained in SSF (compare inset in Fig. \ref{fig5}(g) with Fig. \ref{fig2}(d)). Another exotic spin texture is discovered for finite $h_z$, starting from either SQ state or the ASkL state (see Fig. \ref{fig5}(f)). The skyrmion density map shows clear diagonal stripes of modulating positive and negative values (see Fig. \ref{fig5}(h)). The $\chi$SF (inset in Fig. \ref{fig5}(h)) confirms this phase as a SkDW state where the dominant $\chi$SF peak is located neither at ($0,0$) nor at ($\pi$,$\pi$), but at a general ($Q,Q$) point. 

\noindent
{\it Conclusion:--} 
We derive a generalized CSE Hamiltonian starting with a half-filled Rashba coupled FKLM. Our derivation uncovers 
an elegant analogy with the effective quantum spin model originating from the Mott insulator \cite{Farrell2014a, Banerjee2014}, leading to a general conclusion that Hund's coupling can serve as a proxy for Hubbard interactions for the purpose of understanding the influence of SOC on magnetic ordering. Monte Carlo simulations on the CSE model uncover a rich variety of magnetic states with intriguing textures. The most notable ones are, (i) a liquid-like AFS state with freely orienting AFM stripes, (ii) a state hosting isolated ASk quasiparicles, and (iii) a novel SkDW that generalizes the concept of ferromagnetic and antiferromagnetic skyrmion lattices. 
Since the key ingredients in the microscopic model are the Hund's coupling and the Rashba SOC, 
the new magnetic phases can be realized in thin films of multi-orbital systems with $4d$ and $5d$ elements. More specifically, thin films of Ca$_2$RuO$_4$, SrTcO$_3$ and NaOsO$_3$ have the ingredients to display the magnetic states theoretically reported here
\cite{Sutter2017, Mravlje2012, Rodriguez2011, Franchini2011, Middey2014}. Given the lack of sensitivity of large spins to quantum fluctuations, Hund's coupled multi-orbital systems are strong candidates for hosting ASk quasiparicles. Our work, therefore, provides a guiding principle for stabilizing isolated ASk textures for applications in spintronics.

\noindent
{\it Acknowledgments:}
We acknowledge the use of computing facility at IISER Mohali.

\bibliographystyle{apsrev4-1} 
%\bibliography{Rashba-DE,Spin-orbit-models,AF-Skx}
%

\newpage

\setcounter{equation}{0}
\setcounter{figure}{0}

\renewcommand{\thefigure}{S\arabic{figure}}
\renewcommand{\theequation}{S\arabic{equation}} 

\onecolumngrid
\begin{center}
{\bf {\Large{Supplemental Material}}}
\end{center}

\section{Derivation of the Classical superexchange Model}
The ferromagnetic Kondo lattice model (FKLM) in the presence of Rashba coupling on a square lattice is described by the Hamiltonian,

\begin{eqnarray} \label{eq:KLM}
H & = & - t \sum_{\langle ij \rangle,\sigma} (c^\dagger_{i\sigma} c^{}_{j\sigma} + \text{H.c.}) 
+ \lambda \sum_{i} [(c^{\dagger}_{i \downarrow} c^{}_{i+x\uparrow} - c^{\dagger}_{i\uparrow} c^{}_{i+x\downarrow}) \nonumber \\
& & + \textrm{i} (c^{\dagger}_{i\downarrow} c_{i+y\uparrow}+c^{\dagger}_{i\uparrow} c^{}_{i+y\downarrow}) + \text{H.c.}] - J_\text{H} \sum_{i} {\bf S}_i . {\bf s}_i,
\end{eqnarray}

\noindent
where all symbols have the same meaning as in the main text. For large $J_\text{H}$, it is useful to work in a site-dependent spin-quantization basis instead of a global up-down basis. This change of basis is achieved 
via local $SU(2)$ rotations of the quantization axis, given by,

\begin{center}
	$\begin{bmatrix}
	c_{i\uparrow} \\
	c_{i\downarrow}
	\end{bmatrix} 
	=
	\begin{bmatrix}
	\cos(\frac {\theta_i}{2})   & -\sin(\frac {\theta_i}{2}) e^{-\textrm{i} \phi_i} \\
	\sin(\frac {\theta_i}{2}) e^{\textrm{i} \phi_i} & \cos(\frac {\theta_i}{2}) 
	
	\end{bmatrix}  \begin{bmatrix}
	d_{ip} \\
	d_{ia}
	\end{bmatrix}$.
\end{center}

\noindent
Here, $d_{ip} (d_{ia})$  annihilates an electron at site ${i}$ with spin parallel (anti-parallel) to the localized spin. The polar and azimuthal angle pair \{$\theta_i, \phi_i$\} specifies the orientation in three dimensions of the local moment ${\bf S}_i$. 
% It is easy to see that in the rotated frame the Hund's coupling term contains only the local $z$-components.
% However, this simplification in the coupling term comes at the cost of a complicated hopping structure.

The transformed Hamiltonian takes the form,

\begin{equation}
 H = \sum_{\langle ij \rangle \sigma \sigma^{\prime}} \big( g_{ij}^{\sigma \sigma^{\prime}} d_{i \sigma}^{\dagger} d_{j \sigma^{\prime}} + \text{H.c.} \big) - \frac{J_\text{H}}{2} \sum_i \big(n_{ip} - n_{ia} \big),
\end{equation}
\noindent
where $\sigma \in \{p,a\}$. The advantage of transformation to local basis is that the coupling term takes a diagonal form. This, however, comes at a cost that the spin dependence enters the hopping parameters. The projected hopping amplitudes $ g_{ij,\gamma}^{\sigma \sigma^{\prime}} = t_{ij,\gamma}^{\sigma \sigma^{\prime}} + \lambda_{ij,\gamma}^{\sigma \sigma^{\prime}} $ have contributions from standard tight-binding hopping integral $t$ and the Rashba spin-orbit coupling $\lambda$, where site $j = i + \gamma$ is the nn of site $i$ along spatial direction $\gamma \in \{x,y\}$.

\pagebreak

\noindent
The parallel to parallel components, $g_{ij}^{pp}$, are given as,

\begin{eqnarray}
 t_{ij,\gamma}^{pp} &=& -t  \Big[\cos(\frac {\theta_i}{2}) \cos(\frac {\theta_j}{2}) + \sin(\frac {\theta_i}{2})  \sin(\frac {\theta_j}{2}) e^{-\textrm{i} (\phi_i - \phi_j)}  \Big] \nonumber \\
 \lambda_{ij,x}^{pp} &=& \lambda \Big[\sin(\frac {\theta_i}{2}) \cos(\frac {\theta_j}{2}) e^{-\textrm{i} \phi_i} - \cos(\frac {\theta_i}{2})  \sin(\frac {\theta_j}{2}) e^{\textrm{i} \phi_j}  \Big] \nonumber \\
 \lambda_{ij,y}^{pp} &=& \textrm{i} \lambda \Big[\sin(\frac {\theta_i}{2}) \cos(\frac {\theta_j}{2}) e^{-\textrm{i} \phi_i} + \cos(\frac {\theta_i}{2})  \sin(\frac {\theta_j}{2}) e^{\textrm{i} \phi_j}  \Big]
\end{eqnarray}

\noindent
The parallel to anti-parallel components, $g_{ij}^{pa}$, are given as,

\begin{eqnarray} \label{eq:gpa}
 t_{ij,\gamma}^{pa} &=& -t \Big[-\cos(\frac {\theta_i}{2}) \sin(\frac {\theta_j}{2}) e^{-\textrm{i} \phi_j} + \sin(\frac {\theta_i}{2})  \cos(\frac {\theta_j}{2}) e^{-\textrm{i} \phi_i}  \Big] \nonumber \\
 \lambda_{ij,x}^{pa} &=& -\lambda \Big[\cos(\frac {\theta_i}{2}) \cos(\frac {\theta_j}{2}) + \sin(\frac {\theta_i}{2})  \sin(\frac {\theta_j}{2}) e^{-\textrm{i} (\phi_i+\phi_j)}  \Big] \nonumber \\
 \lambda_{ij,y}^{pa} &=& \textrm{i} \lambda \Big[\cos(\frac {\theta_i}{2}) \cos(\frac {\theta_j}{2}) - \sin(\frac {\theta_i}{2})  \sin(\frac {\theta_j}{2}) e^{-\textrm{i} (\phi_i+\phi_j)}  \Big]
\end{eqnarray}

\noindent
The anti-parallel to parallel components, $g_{ij}^{ap}$, are given as,

\begin{eqnarray} \label{eq:gap}
 t_{ij,\gamma}^{ap} &=& -t \Big[\cos(\frac {\theta_i}{2}) \sin(\frac {\theta_j}{2}) e^{\textrm{i} \phi_j} - \sin(\frac {\theta_i}{2})  \cos(\frac {\theta_j}{2}) e^{\textrm{i} \phi_i}  \Big] \nonumber \\
 \lambda_{ij,x}^{ap} &=& \lambda \Big[\cos(\frac {\theta_i}{2}) \cos(\frac {\theta_j}{2}) + \sin(\frac {\theta_i}{2})  \sin(\frac {\theta_j}{2}) e^{\textrm{i} (\phi_i+\phi_j)}  \Big] \nonumber \\
 \lambda_{ij,y}^{ap} &=& \textrm{i} \lambda \Big[\cos(\frac {\theta_i}{2}) \cos(\frac {\theta_j}{2}) - \sin(\frac {\theta_i}{2})  \sin(\frac {\theta_j}{2}) e^{\textrm{i} (\phi_i+\phi_j)}  \Big]
\end{eqnarray}

\noindent
The anti-parallel to anti-parallel components, $g_{ij}^{aa}$, are given as,

\begin{eqnarray}
 t_{ij,\gamma}^{aa} &=& -t  \Big[\cos(\frac {\theta_i}{2}) \cos(\frac {\theta_j}{2}) + \sin(\frac {\theta_i}{2})  \sin(\frac {\theta_j}{2}) e^{\textrm{i} (\phi_i - \phi_j)}  \Big] \nonumber \\
 \lambda_{ij,x}^{aa} &=& \lambda \Big[\sin(\frac {\theta_i}{2}) \cos(\frac {\theta_j}{2}) e^{\textrm{i} \phi_i} - \cos(\frac {\theta_i}{2})  \sin(\frac {\theta_j}{2}) e^{-\textrm{i} \phi_j}  \Big] \nonumber \\
 \lambda_{ij,y}^{aa} &=& \textrm{i} \lambda \Big[-\sin(\frac {\theta_i}{2}) \cos(\frac {\theta_j}{2}) e^{\textrm{i} \phi_i} - \cos(\frac {\theta_i}{2})  \sin(\frac {\theta_j}{2}) e^{-\textrm{i} \phi_j}  \Big]
\end{eqnarray}

\pagebreak

For large $J_\text{H}$, it is easy to see that at half filling (one electron per site) the ground state corresponds to an insulator, and the situation is similar to the Mott state in the Hubbard model. Therefore, in order to derive an effective spin Hamiltonian it is sufficient to consider a pair of sites. Hence, for the next part the summation over $\langle ij \rangle$ is omitted and we focus on a perturbative treatment of a two-site problem. 

\begin{eqnarray} \label{eq:transfomed KLM}
H & = & H^{\prime} + H_0 \\
  \nonumber
  & = & \sum_{\sigma \sigma^{\prime}} (g_{12}^{\sigma \sigma^{\prime}} d^\dagger_{1\sigma} d^{}_{2\sigma^{\prime}} + \text{H.c.}) - \frac{J_\text{H}}{2} \sum_{i} (n_{ip}-n_{ia})  ,
\end{eqnarray}

\begin{figure}[h] \label{smfig1:exchange_schematic}
\centering
\includegraphics[width=0.6 \columnwidth,angle=0,clip=true]{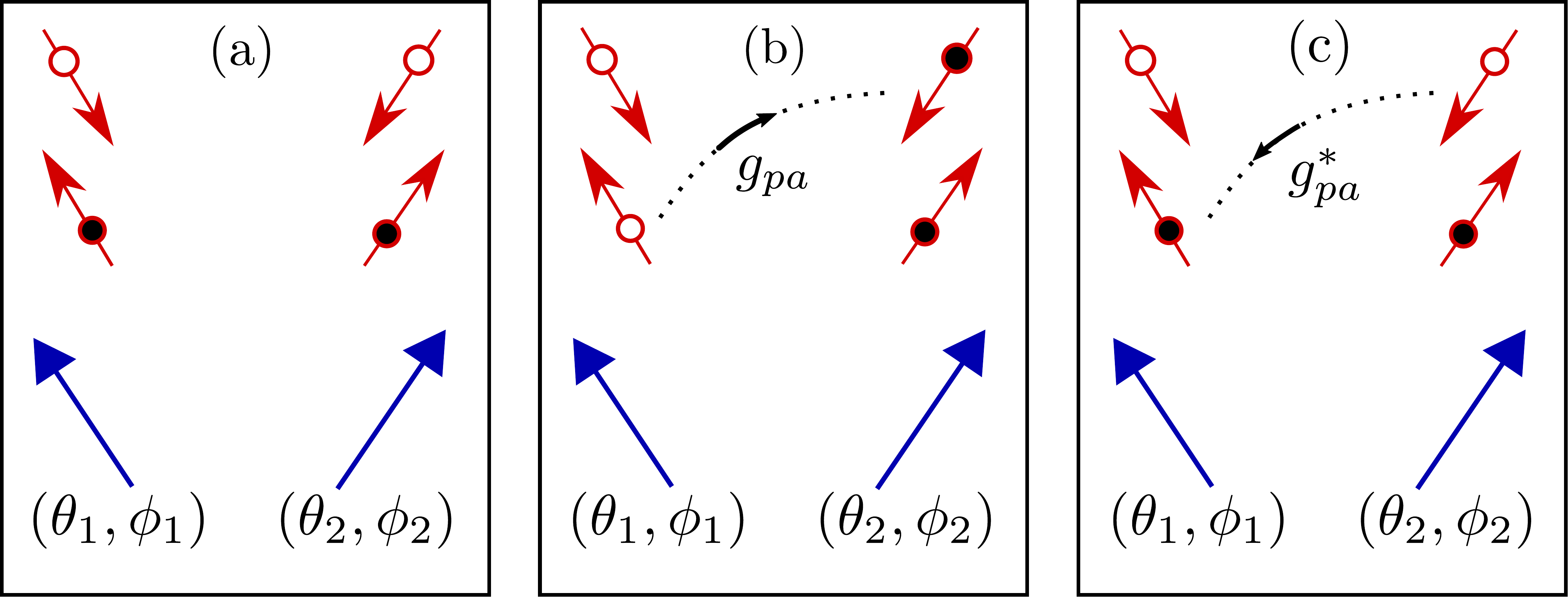}
\caption{Schematic diagram representing one of the virtual hopping process (parallel to anti-parallel) as a pictorial equivalent of the $2^{\text{nd}}$ order perturbation theory.}
\end{figure}

\noindent
In the large $J_\text{H}$ limit we can treat the $1^{\text{st}}$ term, $H^{\prime}$, as perturbation on $H_0$. The ground state of the unperturbed Hamiltonian, $H_0 = - \frac{J_\text{H}}{2} \sum_{i} (n_{ip}-n_{ia})$ in Eq. (\ref{eq:transfomed KLM}), can be written in $2^{\text{nd}}$ quantization notation as,

\begin{eqnarray}
\nonumber
\ket{\psi_0} = d^\dagger_{2p} d^\dagger_{1p} \ket{0}
\end{eqnarray}

\noindent
where $\ket{0}$ denotes the vacuum state. The $1^{\text{st}}$ order correction to energy turns out to be zero:
\begin{eqnarray}
\Delta E^{(1)}_\mathbf{\{S\}} & = & \braket{\psi_0|H_p|\psi_0} \\
        \nonumber
        & = & \braket{0 | d_{1p}d_{2p} (\sum_{\sigma \sigma^{\prime}} (g_{12}^{\sigma \sigma^{\prime}} d^\dagger_{1\sigma} d^{}_{2\sigma^{\prime}}+H.c) d^\dagger_{2p}d^\dagger_{1p} | 0 } \\
        \nonumber
        & \equiv & 0 .
\end{eqnarray}

\noindent
The second order correction is calculated as,

\begin{eqnarray} \label{eq: 2nd order correction}
\Delta E^{(2)}_\mathbf{\{S\}} = \mathlarger{\mathlarger{\sum}}_k \frac{ |\braket{\psi_k|H^{\prime}|\psi_0}|^2 }{ E_0-E_k } = -\Bigg[ \frac{|g_{12}^{pa}|^2}{J_\text{H}} + \frac{|g_{12}^{ap}|^2}{J_\text{H}} \Bigg]
\end{eqnarray}

\noindent
% where it is easy to convince that $E_0-E_k=-J_H$ and the excited states $\ket{\psi_k}$'s are $d^\dagger_{p} d^\dagger_{a} \ket{0}$ and $d^\dagger_{a} d^\dagger_{p} \ket{0}$

\noindent
% We can expand the term $ \braket{0 | d_{p}d_{a} (\textstyle \sum_{\sigma \sigma^{\prime}} (g^{\sigma \sigma^{\prime}} d^\dagger_{\sigma} d^{}_{\sigma^{\prime}}) d^\dagger_{p}d^\dagger_{p} | 0 } $ and using Wick's contraction of fermion we see that only the parallel to anti-parallel hopping term ($g^{ap}$) term survives
% 
% Considering all the excited states, we can rewrite the $2^{\text{nd}}$ order energy correction term of Eq. (\ref{eq: 2nd order correction_1}) as,
% \begin{eqnarray} \label{eq: 2nd order correction_2}
% \Delta E^{(2)}_\mathbf{\{S\}} = -\Bigg[ \frac{|g_{pa}|^2}{J_K} + \frac{|g_{ap}|^2}{J_K} \Bigg]
% \end{eqnarray}

Using Eq. (\ref{eq:gpa}) and Eq. (\ref{eq:gap}) we find, for a general nearest neighbour pair $\langle ij \rangle$,

\begin{eqnarray} \label{eq:SH}
|g^{pa}_{ij,x}|^2 = |g^{ap}_{ij,x}|^2 &=&  \frac {1}{2} \bigg[ t^2(1-{\bf S}_i \cdot {\bf S}_j) + \lambda^2(1+{\bf S}_i \cdot {\bf S}_j-2S_i^{y}S_j^{y}) - 2t\lambda({\bf S}_i \times{\bf S}_j)_y \bigg], \nonumber \\
|g^{pa}_{ij,y}|^2 = |g^{ap}_{ij,y}|^2 &=& \frac{1}{2} \bigg[t^2(1-{\bf S}_i \cdot {\bf S}_j) + \lambda^2(1+{\bf S}_i \cdot {\bf S}_j-2S_i^{x}S_j^{x}) + 2t\lambda({\bf S}_i \times{\bf S}_j)_x \bigg].
\end{eqnarray}
\noindent
Putting back the expression from Eq. (\ref{eq:SH}) in Eq. (\ref{eq: 2nd order correction}) and bringing back the sum over sites, we arrive at the CSE model defined in the Eq. (2) of the main text.

While the similarity between the above derivation and that for the Hubbard model is already clear, we emphasize on a few important differences.

\begin{enumerate}
 \item The derived model is explicitly classical since the local moment variables were assumed classical in starting FKLM.
 \item The starting model is a parameterized single-particle model, as opposed a truly many-body Hubbard model. parameterized single particle means, for given classical spin configuration the quantum problem is a single-particle problem.
 \item A consequence of the point 2 above is that the $2^{\text{nd}}$ order perturbation theory presented above is rigorous. For the Hubbard model, an accurate derivation is rather involved and requires Schrieffer–Wolff transformation.
\end{enumerate}

Despite the above differences, what is remarkable is that by promoting the classical spin variable to operators the model becomes identical to the generalized Heisenberg model obtained from the Hubbard model.

\section{Energy comparison}

\begin{figure}[H]
\centering
\includegraphics[width=0.6 \columnwidth,angle=0,clip=true]{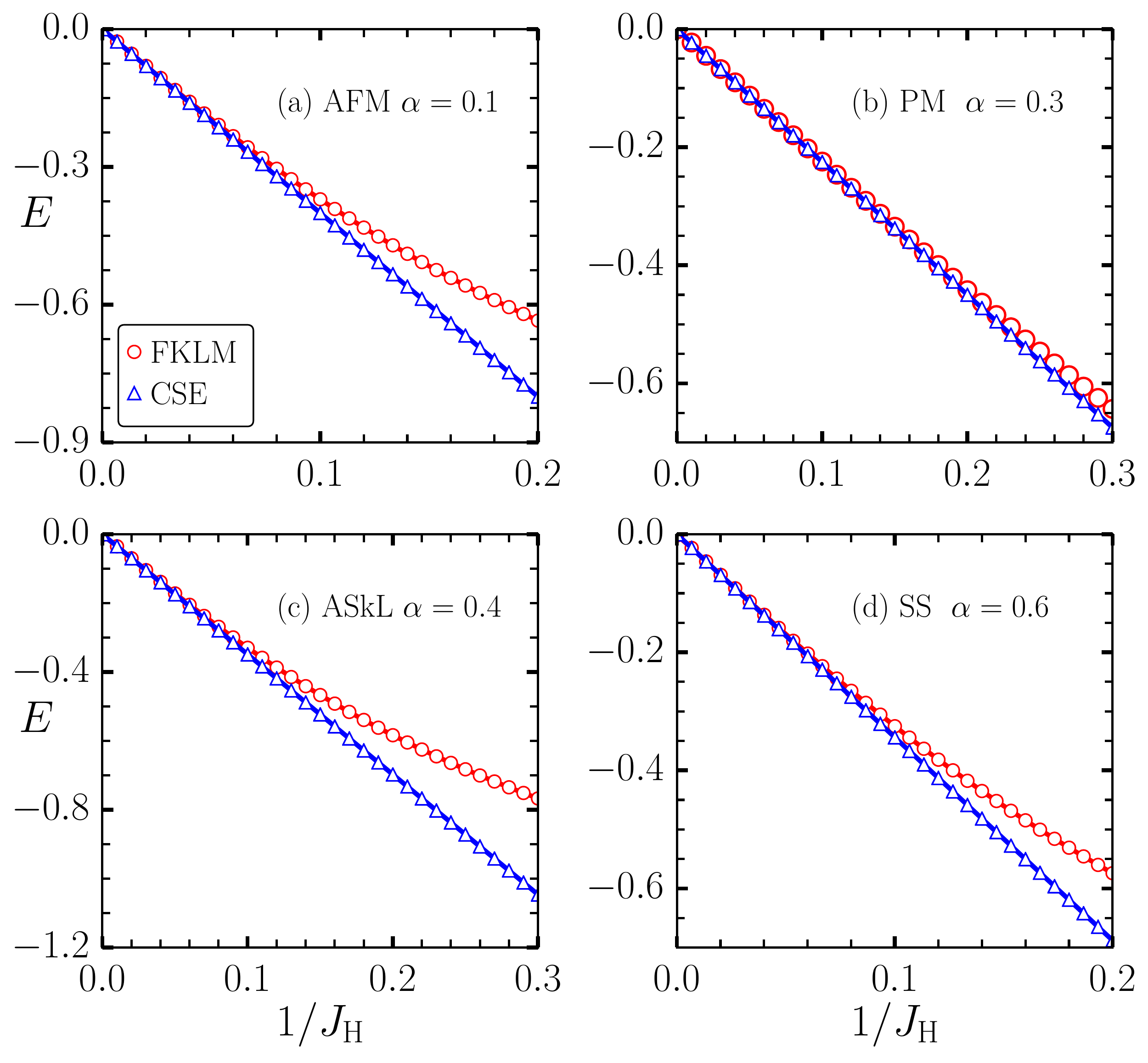}
\caption{$E$ vs. $1/J_\text{H}$ comparison between ferromagnetic Kondo lattice model (FKLM) and classical superexchange (CSE) model. The different spin states are AFM, PM, ASkL and SS for (a) $\alpha=0.1$, (b) $\alpha=0.3$, (c) $\alpha=0.4$, (d) $\alpha=0.6$ respectively.}
\label{smfig2:energy_compare}
\end{figure}

In order to test the validity of the derived CSE model, we directly compare the energies of different spin configurations with those obtained in the spin-fermion model Eq. (\ref{eq:KLM}). We show the energy comparison as a function of $1/J_\text{H}$ (see Fig. \ref{smfig2:energy_compare}) for different spin configuration at representative $\alpha$ values. We can see from the comparison, the energies matches quite well up to $1/J_\text{H} = 0.2$.

\noindent

% \section{Confirming degenerate spiral states}

\section{Origin of antiferromagnet string (AFS) state}

We find an exotic antiferromagnetic string (AFS) state to be stable in range, $0.06 \leq \alpha \leq 0.38$ as a parent state of sASk state. Here we show that the inhomogeneous AFS states are a consequence of pseudo-dipolar interaction term in the Hamiltonian. For small values of $\alpha$, terms proportional to $\lambda^2$ in the Hamiltonian Eq. (\ref{eq:SH}) can be ignored in comparison to $t\lambda$ term. The $t\lambda$ terms prefer spiral states with a peculiar degeneracy that we show below. It is clear from the form of the Hamiltonian that, along $x$-direction a spiral in $xz$ plane is preferred and a spiral in $yz$ plane is preferred along $y$-direction.

We construct a simple variational ansatz to study the degeneracy of general spiral states. The following variational spin configurations allow the plane of the spiral to change via parameter $\Phi_p$:

\begin{eqnarray}
S_i^x &=& S_0 \sin(\boldsymbol{q}.\boldsymbol{r}_i)\cos(\Phi_p), \nonumber \\
S_i^y &=& S_0 \sin(\boldsymbol{q}.\boldsymbol{r}_i)\sin(\Phi_p),\nonumber \\
S_i^z &=& S_0 \cos(\boldsymbol{q}.\boldsymbol{r}_i),
\label{eq:vari_ansatz}
\end{eqnarray}
\noindent
where, $S_0=1$ is the magnitude of the classical spin vectors. The orientation of the spiral plane is defined by $\Phi_p$, where $\Phi_p = 0$ represents $xz$ plane spiral and $\Phi_p = \frac{\pi}{2}$ is for $yz$ plane spiral. $\boldsymbol{q} = q(\cos\beta,\sin\beta)$ is the spiral wave-vector.
In the $\alpha$ range $0.06 \leq \alpha \leq 0.38$ , we find that all the spirals are degenerate irrespective of the choice of $\Phi_p$, provided the orientation of the spiral wave-vector is constrained via $\beta-\Phi_p=\pi$. The degeneracy of these spiral states explains filamentary structures in the AFS regime.

Even though the orientations of antiferromagnetic strings are energetically free, change in their width costs energy. This leads to some intermediate level of degeneracy with $\mathcal{O} (e^{\sqrt{N}})$ degenerate configurations, contradictory to a true macroscopic degeneracy with $\mathcal{O} (e^N)$ fold ground states. Monte Carlo update dynamics may get affected by this peculiar behavior of degeneracy while exploring the configuration space.

% In order to quantify this degeneracy of spiral states, we define $\Delta E = \max[E_{min}(\Phi_p)] - \min[E_{min}(\Phi_p)]$. Exact degeneracy of states exists if $\Delta E = 0$. We show the variation of $\Delta E$ with the coupling constant $\alpha$ as an inset in Fig. \ref{fig:vari_E vs beta} (b). The degree of degeneracy clearly reduces near $\alpha=0.35$, which coincides with the crossover point between CSL and SQ spiral states.

\begin{figure}[h]
    \includegraphics[width=0.6 \columnwidth,angle=0,clip=true]{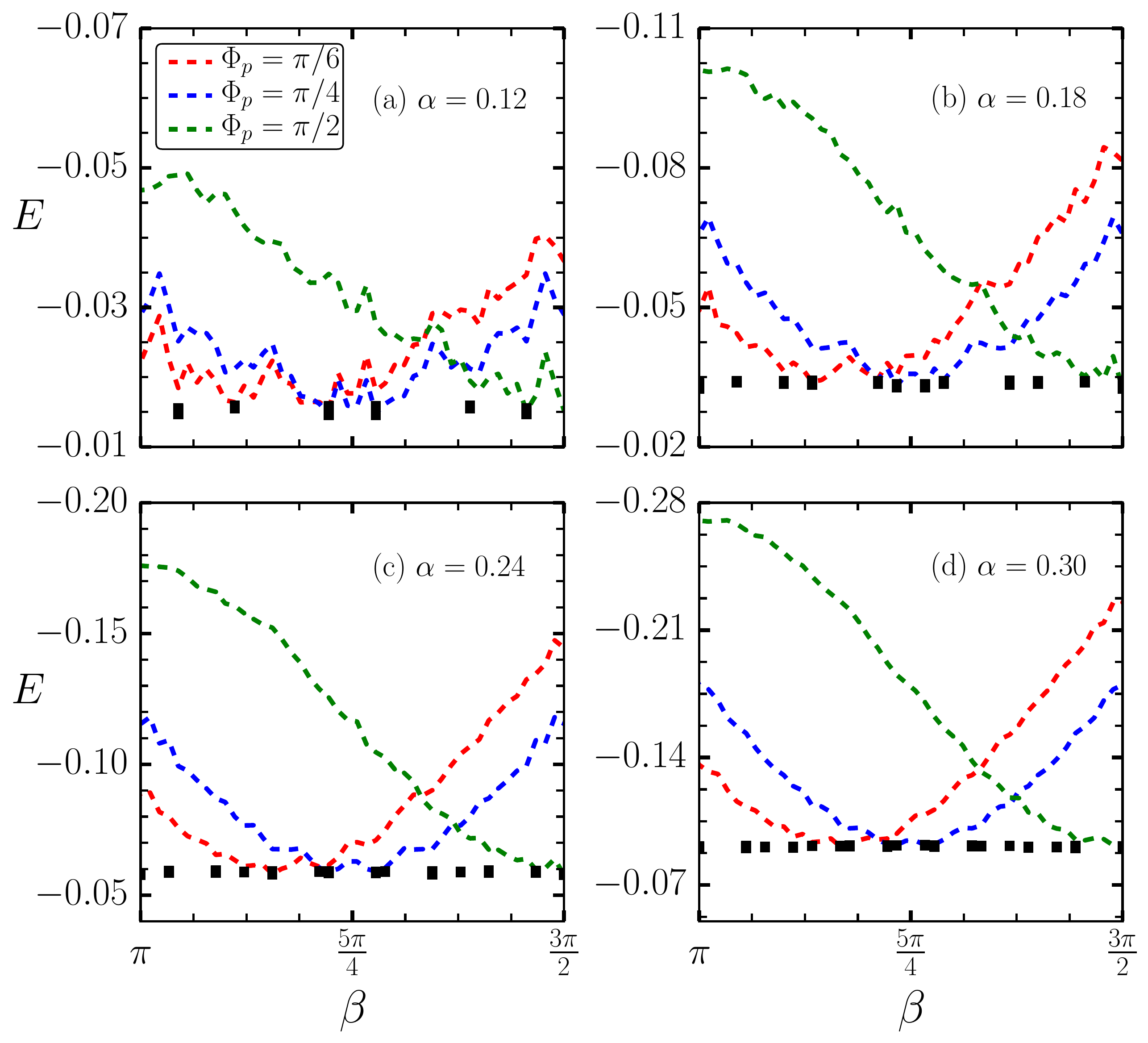}
    \caption{Energy per site $E$ as a function of wave-vector direction $\beta$, for 
            (a) $\alpha$ = 0.12,
            (b) $\alpha$ = 0.18,
            (c) $\alpha$ = 0.24 and
            (d) $\alpha$ = 0.30,
            obtained for states defined via variational ansatz Eq.(\ref{eq:vari_ansatz}). Energy is minimized over the magnitude $q$ of $\boldsymbol{q}$. The filled squares mark the value of $\beta$ that corresponds to the minimum energy for a given $\Phi_p$. Almost identical values of this minimum energy for different $\Phi_p$ confirm the unusual degeneracy that exists in the AFS regime.
            }
    \label{smfig3:vari_E vs beta}
\end{figure}

\noindent

\end{document}